Mr. Moo's First RPG: Rules, Discussion and the Instructional Implications of Collective Intelligence on the Open Web

Imagine an active online learning community of writers, artists, and designers, many spending more than eight hours a week composing projects. In this community, young people, primarily between the ages of 18-26, regularly critique, facilitate, and support each other in their composition activities. They are motivated to participate by their shared interest in their creative work. In the age of Wikipedia, this might not seem particularly novel, but what I am describing is an online discussion board, RPGmakerVX.net. Elsewhere, I have presented a general outline of the kinds of individuals involved in this community and the way that the site as a whole functions as an open learning environment (Owens, 2010). In this essay, I present a case study of one participant in this community. His username is Mr. Moo, and at the time I interviewed him, he was a 19 year old college student from Calgary, Canada. He created his first role-playing game, *Prelude of Identity,* when he was eighteen.

After providing a conceptual context for this case study on collective intelligence, I suggest that the discussion board rules and interaction enable a dialogue around composition that gives Mr. Moo a valuable learning experience while producing a role-playing game. The perspective of collective intelligence enables educators to use interest-driven, online communities as open education tools in more formal learning environments.

**Formal and Informal Learning on the Web**

Along with enabling discussion, web forums are searchable repositories of knowledge. Pierre Levy (1999) has conceptualized these kinds of discursive spaces as a form of collective intelligence. In this view, the web contains a shared pool of knowledge that is collectively produced and consumed.

For example, in a study of discussion threads in the *World of Warcraft* forums, Steinkuehler and Duncan (2008) found that beyond serving as a space for discussion, the threads also functioned as a knowledge base. Steinkuehler and Duncan suggest that the collaborative construction of knowledge in *Warcraft* forums parallels the kind of collaborative construction of knowledge that occurs in scientific communities, indicating the sophistication of these players' arguments.

Levy's concept of collective intelligence proposes that readers, spectators, producers, creators, and their respective interpretations are blending into a reading-writing continuum. Levy suggests that this continuum "will extend from the machine and network designers to the ultimate recipient, each helping to sustain the activities of the others" (p. 28). From this perspective, the product (the game), the process (discussion on the boards), and the learning are all part of a distributed network in the reading-writing continuum.

We can see another component of this reading-writing continuum in the way players are invited to modify games. Squire and Giovanetto (2008) suggest that the *Civilization* forums act as scaffolding, enabling gamers to develop the ability to modify the game. Gamers participating in these forums clearly develop technical skills. Beyond that, however, Squire and Giovanetto (2008) argue "More important than the particular facts or technical processes may be the practice of negotiating social organizations (including forming them) to further one's own learning" (p. 27). As these informal online learning spaces become more commonplace, the ability to navigate such spaces is becoming an increasingly important skill.

Similar patterns of learning emerge in online communities based on *The Sims*. Gee and Hayes (2010) have illustrated how gamers who write fanfiction and create machinima developed as creators and found their voices through sharing their work in online discussion boards. For Gee and Hayes, there is a stark contrast between the kinds of learning that occurs in online communities and learning in schools. They suggest that schools, "which now stand so separate from the rest of the learning landscape, will have to integrate with other means and locations of learning" (p. 150).

Together these examples illustrate various kinds of media production skills, what literacy specialists Colin Lankshear and Michele Knobel (2003) call new literacies. Lankshear and Knobel have encouraged scholars to explore those communities to develop innovative ideas for formal learning environments. More recently, Greenhow, Robelia, and Hughes (2010) issued a call for exploring learning in informal online learning communities as a means to invigorate classroom practice.

Concurrent with this move to study informal leaning communities is a substantive attempt to "open" formal education. Often described as the "open education movement," the goal here has been to expose course content, primarily in higher education, to anyone online. For example, projects like MIT's open courseware and broader initiatives like the Open Educational Resources initiative serve this goal. John Sealy Brown (2008) has suggested that increasingly available open courseware, access to powerful instruments and simulation models, and scholarly websites are making the resources of the academy accessible to anyone.

To some extent, this model of openly exposing educational resources and enabling self-education is strikingly similar to ideas from Ivan Illich. Illich (1971) suggested that the institution of school itself was the central problem of education. He argued that a new system should be created, one that would "depend on self-motivated learning instead of employing teachers to bribe or compel the student to find the time and the will to learn." Furthermore, such a system could "provide the learner with new links to the world instead of continuing to funnel all educational programs through the teacher" (p. 73). He referred to this potential network of learners as an "opportunity web" or "learning web," the central idea being that a network of peers and elders in any range of subjects could facilitate learning.

The open education movement often uses Illich's work as a model for thinking about its work as learning webs (Peters 2008; Leinonen, Vadén, & Suoranta, 2009). At the same time, work on online gaming affinity spaces exemplifies many of the characteristics of learning webs as open, interest-driven spaces. Acknowledging collective intelligence as a framework for understanding affinity spaces provides the potential to think in more nuanced ways about how educators might use these online affinity sites as teaching tools.

In both affinity communities and the open education movement, it is unclear what role professionals play. At the same time, as Squire suggested, the fact that anyone can find these online communities does not mean that people are even aware of their possibilities. There is still a clear need for the equity-producing capabilities of educational communities. Illich's idea of the learning web still requires guides to make use of these networks. All of this content—the individuals, the discussions—can be thought of as part of our common collective intelligence.

To provide a model for how affinity communities function and how collective intelligence is enacted, I present a case study of Mr. Moo, tracing his interactions with individuals on the RPGmakerVX.net discussion boards. Briefly, RPGmakerVX.net is a discussion forum where users of the role-playing game building software, *RPG Maker VX*, discuss and critique game designs, artwork, music, and computer code. I examine how the discussion board site, the discussion board rules, Mr. Moo, and the other forum participants create a valuable learning experience for those involved. By examining the rules of the community, Mr. Moo's posts, the responses to them, and his reflections on them, I illustrate how this learning space embodies collective intelligence.

## Discussion Board Rules

Like most discussion boards, RPGmakerVX.net has an explicit set of rules. There are a series of generic governing rules that apply to all parts of the discussion board and more specific rules that apply to certain kinds of discussions. These board rules speak to some of the governing values of the community.

### Defining Roles in the Board Rules

The board rules start by identifying "The Prime Directives," the "rules which override the other rules in most cases." The first of these directives explains, "Elitist bastards and elitist bastardy behavior will not be tolerated." "Elitist bastards" are defined in part by their refusal "to learn the distinction between "newb" and "n00b." The differences between these ideas help shed light on some of the social mechanisms that undergird the idea of collective intelligence. The newb is inexperienced, but wants to learn, and when given guidance is happy to take and act on it. In contrast, the n00b, while similarly clueless, is unwilling to accept that he or she should take guidance from those with more experience. RPGmakerVX.net has almost no barrier to entry. All anyone needs to do is sign up for an account to join and start posting. This means that new community members are vetted after they have already come in the virtual door. The central rule for the site exists to ensure that experienced community members nurture participation from new members who seek guidance. In terms of collective intelligence, these terms define the roles for teacher and learner in an open knowledge creation space.

**The Rules for Posting Completed and Underdevelopment Games**

The rules for the "completed games" and "early project feedback discussion" have a stern tone to them. If participants want to create a new thread about their games, they "are expected to read, understand the following rule and meet the minimum requirement to post a new thread in this forum." The requirements for the game posts speak to what the community values and how it organizes discourse. Specifically, the focus on making sure that the posted games are free of bugs, posts include substantive text related to the storytelling components (setting, plot, characters) of their games, and posts include a set of screenshots from the game that clearly demonstrate the designer's skills at mapping (creating game maps) and eventing (a way to set triggers and conditions for events to unfold in the game).

The rules request significant information about a game's story. These include a story synopsis of at least 350 words that "needs to cover how your story will flow together, what the character's basic interests in the story are." There must be at least 225 words on world development. The stern tone of these rules continues in the following imperative: "If you can't write down at least at least 225 words then you may want to rethink your game idea." Posters must share character bios of at least one hundred words for all playable characters and their "characters need to be developed beyond just their archetype and basic personality." The focus on storytelling components helps would-be game players make decisions about playing the game. They also make it easier for potential players, and other site community members, to critique the games' stories.

The mapping requirements ask for a minimum of 12 screenshots. These screenshots "must show more than just the basic and generic town." These rules are supposed to show "mapping skills, eventing skills, and NPC's." The screenshots are intended to provide a view into one's skills at constructing a game.

These rules for starting a new game thread are substantial so that as little of an imposition is made on potential players and game reviewers as possible. These rules create a single document that exhibits story and game design components in a format that others can easily digest and respond to. These rules also create a first post which is routinely updated with additional revised content, acting as a kind of homepage for the game.

The word counts for each section are in bold, providing emphasis on clearly measurable requirements. While I realize that word length is not a proxy for quality writing, it is what this community has focused on, and reporting it offers a sense of the shape of community members' discursive interactions.

At points, the stern tone of the posting rules takes on an air of condescension. Readers can almost hear how tired the moderators who wrote these rules are of seeing people fail to post substantive detail. Each poorly formed post forces moderators to spend more of their time bugging posters to share the information that commenters need to provide substantive feedback.

### Discussing Mr. Moo's Game on the Boards

In April of 2009, Mr. Moo started a discussion thread to share a full draft of his game, *Prelude of Identity*. The discussion that follows his post provides an opportunity to explore how

RPGmakerVX.net users interact with each other, enact the discussion board rules, and create a learning environment for Mr. Moo.

    Mr. Moo's initial post is just over 1,400 words long. The content of this post underscores the significant amount of work that goes into presenting a game, provides a flavor for how the game posts works, and illustrates what successful interpretation of the rules looks like. Mr. Moo presents just fewer than three hundred words explaining a story of knights and political intrigue. He provides slightly more than three hundred words on the setting; the story takes place in fictional European country, and he describes how the class system works in this country and its relationship to other European powers. The post then includes 550 words about the three different characters. Each bio explains a bit about the characters' backstories and motivations and also includes information about how he has customized their capabilities in the game's combat system. In the next one hundred words, he provides a listing of the nine different scripts he used (*RPG Maker VX* has a Ruby-based scripting language that allows players to significantly alter the functionality of the software), credits for the music and artwork he used, the number of quests available in the game (21), and the total number of maps (165). The remainder of the post includes 16 screenshots, a link to download the game, and a link to a review of the game from the primary administrator of RPGMakerVX.net. All together, the information provided gives any potential player a quick sense of the story, artwork, and gameplay. He has demonstrated that he can follow the rules and meet the requirements for the game post. While new to the discussion boards, he is clearly not a N00b.

    The first two responses to Mr. Moo's post are procedural questions. The first poster questions if he has included all the relevant information. Mr. Moo linked to a different RPG Maker discussion board where he had posted developmental versions of the game, and this

poster expressed concern that Mr. Moo hadn't sufficiently moved all of the relevant content from that other site to his post on RPGMakerVX.net. Mr. Moo responds to this question by explaining that he already had moved everything over, and in keeping with the rules for posting a completed game, this post had more images of the game than any other place he had shared information about his work. An administrator of the site then notes that the moderators had reviewed the post and that it met the requirements for it to be moved into the completed games discussion section.

The rest of the discussion thread involves exchanges of feedback about the technical and compositional components of the game, starting with a mixture of requests for gameplay help, along with criticism and praise of the game's story, style, and technical features. Altogether more than 60 players provide comments and criticism. Collectively, this discussion illustrates how extensive an opportunity for feedback the RPGmakerVX.net site can be for someone who wants to develop as a game designer. The following comment is indicative of several of the important themes that emerge from this commentary. "OH MY GOODNESS THE ENCOUNTER RATE IN THE FIRST CITY AT NIGHT IS EXCRUTIATING. Just so you know. I'm really enjoying this game; the storyline is very enjoyable, and I like that the gameplay is different." The commenter starts with an all caps exclamation about the encounter rate, the frequency that random monsters attack the player in the game. By the third sentence, the commenter switches to praise, offering a generally positive comment and drawing attention to the fact that the commenter enjoys the story and the gameplay. The positive comments are not particularly specific, but the reviewer wanted to make sure that positive feelings were communicated about the game along with specific criticism.

Throughout the thread, commenters move back and forth between positive and negative comments, but whenever they make the latter, they include hedges such as, "Feel free to ignore my response" or "You deserve praise for the effort put into your game, but obviously it can't be perfect, and that is what I'm focusing on." At several points, commenters explicitly note that they want to offer constructive feedback. The discussion moves from technical issues and bugs, to questions about the author's choice of setting, to comparisons to a series of major role-playing games. While online discourse is often caricatured as antagonistic, the discussions on RPGmakerVX.net quickly demonstrate that this community, like many online communities, prizes decorum and deference. Members' courtesy affirms the importance that they place on the community as a whole and on individual contributions. At each point the commenters are clearly following the rules associated with how to treat new members and avoid being an "elitist bastard."

In the midst of the discussion thread, there are two more substantive reviews of the game that offer feedback and play an important role as part of the collective intelligence enabled through this discussion thread. About two weeks after the game was posted, Pine, one of the forum administrators, posted a 350-word detailed review, and the following day, Hobonicus posted a 1,075-word critique. Pine organizes his comments into a discussion of the game's story, which he found compelling, and playability. After offering praise, and noting that the game has a high difficulty level that might not appeal to some players, the review ends favorably with "Pine's Final Rate" of "8.5/10 stars."

After receiving this rather glowing review from Pine, Mr. Moo updated the initial post to include a direct link to Pine's review, pulling the review out of the threaded discussion and inserting it into what serves as the game's homepage. This is in part only possible because the *Invision Power Board* software that RPGmakerVX.net uses generates anchored links that one can use to link to a particular comment in a given discussion thread. In this sense, this deep linking is a form of collective intelligence. Posts are not strictly to be interacted with as part of a stream of discussion; the software exposes each post as a potential resource that could be linked from and referred to in other contexts. Mr. Moo transformed part of an ongoing discussion into a resource he could use to advertise his game. The structure of Pine's post anticipated this use. Instead of framing it as part of the ongoing discussion, Pine structured the post as a review, complete with headings referring to particular parts of the game he was reviewing and a concluding star rating.

In contrast, Hobonicus gave a review after four hours of play, writing, "I hope I don't come off as harsh, I'm reviewing this directly from what I've played and it's only meant to be constructive." Mr. Moo did not add a link to this review to his original post. Hobonicus discusses some issues he had with the plot; this critique is largely focused on inconsistencies in the character's actions and motivations. He then comments on how he thought the characters weren't particularly interesting, principally the "cut and paste" villain. Hobonicus commends the mapping in the game: in particular, "The towns were very well crafted and the whole world felt alive." The strongest criticism in this review is reserved for problems with the encounter rate noted earlier. "The battles. Oh God the battles. Every. Few. Steps. Another. Fight. Which would be tolerable if the battles were any fun, but they really aren't. They are ridiculously repetitive and boring." To balance the harshness of his critique, Hobonicus ends by pointing out some of the positive points again: "I realize my review has seemed pretty harsh so far, which I didn't intend. There was obviously a lot of effort put into it, which shows." He explains, "I generally sound pessimistic in reviews because I focus on what I think is lacking" and that some "major (and easily doable) tweaks to certain aspects could make this game great." He then highlights several positive points. "I can't stress how awesome the cities are, and how alive the world feels, but I feel the game's progression needs work on several levels." Through all of these comments, Hobonicus offers substantive feedback on the story and the technical components of gameplay. In accordance with the established rules of not being an "elitist bastard," Hobonicus frequently reaffirms respect for Mr. Moo's time and his skills evident in parts of the design. These two reviews are illustrative of the kinds of feedback that anyone who follows the rules of the discussion board can receive.

### Reflection on the Discussion a Year Later

Roughly a year after the release of his game, I was able to interview Mr. Moo over email about his experiences. His recollections offer insight into what he took away from participating in the forums. When asked what kind of feedback he had received on the game from the community, he responded, "Those who send feedback usually criticize the pace, the difficulty, or mapping. Despite this, many say it is one of the more well mapped, well paced games out for VX right now." The feedback he had received clearly helped him refine his craft, and he has used that criticism to inform his next design process. Specifically, he noted, "The biggest change was not to 'write as I work.'" He identified this as a common weakness in his design process that had resulted in the various issues community members reported about his games. For example, he explains how he changed his design approach for his second game, *Crescendo of Identity*, as a result of this experience: "I did the entire database before I started making the game. I would edit the numbers accordingly if I found a player/enemy to strong or too week. Finally I change the pacing of each of the game's segments." Through these changes, he feels he "Took many of the suggestions and simplified the system without removing depth from the game." He had taken the individual criticisms of his work and isolated a common cause in the weakness of his design process, which he explicitly addressed in his development process for his second game.

Mr. Moo also grew as a composer in other ways. Over the course of the thread, he diplomatically defended his work from criticism, responded to substantive critiques, and explained his design decisions. In short, he engaged in the kinds of authentic critical dialogue that designers and writers engage in professionally. While Mr. Moo's experiences are uncommon (most of the participants in the community never finish their games), his experiences and the community's response to those experiences illustrate both how the community encourages reflective creative production through critical dialogue. The rules structure the posts, the posts structure the discussion, the discussion becomes a resource, and ultimately the discussion results in substantive reflection and learning through participation. Together these features serve as an interlocking system that produces knowledge, provides a game, and serves as a tool for teaching through practice.

## What does RPGmakerVX.net Offer Composition Educators?

What does this story about Mr. Moo and the RPGmakerVX discussion boards tell us about games and composition? I think the site, its rules, and the community participating in the site can collectively be used as instructional tools for composition instruction.

The format for writing game proposals is effectively a composition assignment. An instructor could borrow this format as a template for having students think through all of the components and features that they could consider in composing a role-playing game. In this respect, the rules for game proposals represent an assignment "in the wild": a set of requirements for composing a document that have emerged organically from a community of individuals interested in composing games on the *RPG Maker VX* platform. This is particularly helpful as the *RPG Maker VX* software is one of the most inexpensive game composition software packages on the market. The software costs less than sixty dollars and unlike many other platforms there are no licensing fees to release created games.

If one did want to teach composition through this tool, it is worth thinking about how extensive a role consideration of audience plays in the composition of these role-playing games. While all composition involves a consideration of audience, the discussions on RPGmakerVX.net focus our attention on audience members' experiences playing the games. When composing a game, setting the encounter rate for how often players run into monsters involves setting a few variables, while playing through the result of those variables can take a considerable amount of time. The intense community reactions to how these variables are set suggests that if one were to use RPG Maker to teach composition, there would be a need for students to get feedback from potential players. Now, if we are open to considering the rules for proposals as instructional tools, it might also be worth considering the site itself as an instructional tool.

Beyond borrowing the requirements for writing up the proposal, it would be possible to work through how in-class interactions and instructor feedback could be mixed with actually sharing proposals and work with the online community. After giving students feedback on their work and offering them the chance to revise it, instructors could have students post their work on the RPGmakerVX site for further feedback. This could then serve as the basis for class discussion about the kinds of feedback students receive from sharing their work.  The discussion could help students think through how to refine their ideas in the face of external critique. This approach would have the dual benefit of expanding the number of individuals giving feedback on a student's work and also providing an opportunity for instructors to help students think through how to navigate online informal learning communities.

The open educational resources movement suggests that access to educational resources will result in those interested in learning being able to learn as they like. However, the ability to navigate these spaces is becoming an important skill—one that some young people are acquiring and others are not. What is particularly exciting about thinking about using a site like RPGMakerVX.net as an instructional tool is that, aside from having potential benefits for teaching composition, instructor-led interaction with informal online learning communities could help students learn how to navigate other of online spaces in the future.

**Learning to Compose RPGs and Learning to Learn Online**

At this point, what might be the most valuable step for educators is not investing their energy in making the resources they create accessible, but instead in figuring out how their experience, knowledge, and interactions with students can enable us to use the open web of informal learning communities that are springing up around tools like RPGMaker as tools for our own instructional practices. While people like Mr. Moo are left to their own devices to navigate the complex social interactions required for this kind of participation, with only the discussion board rules to guide them, instructors could help students parse the rules for participation, think through how they can best participate in the community, and consider how they should respond to any particularly harsh criticism, all the while developing their composition skills.

Using resources like the *RPGmakerVX.net* discussion board has the potential to create a type of collective intelligence and media production lab. Aside from learning and developing a media production craft, students can learn to navigate and negotiate these kinds of informal online learning environments.